\documentclass[a4paper,12pt,english,german]{article}

\begin{document}

{\bf Which Multiverse?}

\bigskip

M. Dugi\' c and J. Jekni\' c-Dugi\' c

\bigskip

dugic@kg.ac.rs

jjeknic@pmf.ni.ac.rs

\bigskip

{\bf Abstract:} The complex (composite) systems such as the Universe allow the different decompositions into subsystems. The Everett's Many Times Interpretation (MWI) heavily relies on the occurrence of decohernece that should provide the classical reality for the Worlds. However,  applying the occurrence of decoherence as the sufficient condition for the classical reality of the open systems, one seems obliged to consider the different decomposition equally (classically) realistic. But this leads to an inconsistency of the Everett's MWI in its very foundations: why the decomposition as we perceive it should be the only one realistic?

\bigskip

\noindent
The Everett's Many Worlds Interpretation of quantum mechanics is a legitimate and mind-provoking an issue in the foundations of quantum mechanics.
While the consistency of the MWI is an open issue, one can pose another question of fundamental importance not--as to the best of our knowledge--raised in the literature yet. The question we have in mind is both the very heart of this draft, as well as a matter of consistency of MWI in the ontological level.

In this draft, we assume the existence of a noncontroversial, the fully consistent and generally adopted MWI. Particularly, we assume that, somehow, at least the following basic issues of MWI are properly established:

(i) the noncontroversial emergence of the probabilities within the fully unitary evolution of the Multiverse.

(ii) the substantial role of the environment-induced decoherence as the fundamental quantum-mechanical basis for the consistency of MWI.

While we do not predict the final outcome regarding the above points (i) and (ii), we still refer to the point (ii) as raising another point of the fundamental importance for the foundations of MWI.

\bigskip

The necessity of the decoherence effect in the foundations of MWI bears the clear cut physical background as follows: due to the fact that a World
is an {\it isolated} physical whole, there is not the room for the decoherence effect {\bf but}  the division of the World as a pair of "open system + environment ($S+E$)", where $E$ is the internal environment that induces decoherence for $S$. Without any specific physical model for the
composite system $S+E$, we are still able to pose the above-mentioned question as follows in the sequel.

We pose the following question while leaving the technical details for the latter discussion:

{\it (Q1) As every composite system can be nonuniquely decomposed as a collection of the subsystems$^{a}$, we wonder if every possible division of the World into subsystems gives rise to the one and the same conclusion regarding the basic interpretational rule of MWI.}

To this end, in order to be more specific, let us emphasize: the different divisions of the World into subsystems can be mutually related by the linear canonical transformations$^a$ of the basic set of the observables (the position- and the momentum- observables) of the constituent systems. So, given the decomposition $World = S + E$, we have in mind the linear canonical transformations (LCT) of the $S$'s and $E$'s observables in order to redefine the World as: $World = S' + E'$. Given that both decompositions refer to a given instant of time, $t_{\circ}$, our question (Q1) actually poses the question of consistency of MWI regarding the different decompositions, $S+E$ and $S'+E'$, of the one, and in the instant $t_{\circ}$ uniquely defined World.

This question may not be naive as it may seem prima facie.

Let us consider the following elaborations.

(1) Let us first consider the following situation: due to interaction with $S$, the environment $E$ induces decoherence$^b$ for $S$, {\it while} the systems $S'$ and $E'$ mutually do not interact$^c$, and therefore there is not decoherence for the decomposition $S'+E'$. The consequence is then rather obvious: while {\it there is a splitting} of the World regarding the decomposition $S+E$, {\it there is not the splitting} regarding the decomposition $S'+E'$. Certainly, there is a controversy regarding {\it the one and the same World} uniquely defined in the instant $t_{\circ}$.

One possible way out of this controversy may be the following stipulation (implicit to the foundations of the decoherence theory):

(S1) one of the decompositions of the World, either $S+E$ or $S'+E'$ is not physically realistic, and {\it due to the presence of the decoherence effect} in $S+E$, one should consider the later, i.e. the decomposition $S'+E'$, physically {\it artificial (nonrealistic)}.

However, then appears another problem as defined by the following situation.

(2) Let us assume that both decompositions bear decoherence$^d$: $S'$ is (also) decohered by $E'$. In general, one can expect that the two kinds of the
decoherence processes are not mutually equivalent. The rule (S1) sets that both decompositions should be {\it equally realistic}, while, in general {\it not necessarily physically (or information-theoretically) equivalent}$^e$.

Let us be more specific: the different decompositions of state of the World in the instant $t_{\circ}$ read:

$$\sum_i c_i \vert \phi_i\rangle_S \vert \chi_i\rangle_E =
\sum_{\alpha} d_{\alpha} \vert \mu_{\alpha}\rangle_{S'} \vert \nu_{\alpha}\rangle_{E'},  \eqno (*)$$

\noindent
where the subsystems $S$ and $S'$ have nothing in common, while eq.(*) gives for the World in the instant $t_{\circ}$ a nonunique decomposition--{\it the "pointer basis" is not unique}!

This non-equivalence raises the ontological problem, very much like the point (1): which
decomposition into the Everett branches--due to $S+E$, or to $S'+E'$--should be chosen? In other words: which Multiverse should be considered to be physically realistic?

While this question is of the ontological nature (there is not "observer" or any third party in the decomposition of the World in the given instant $t_{\circ}$), we do not find the answer like "both Multiverses  should be regarded realistic" straightforwardly acceptable. In any case, we believe--including the more general situations than those presented by (1) and (2) above--that this issue (the issue of "what is 'system'?"$^e$ in "Technical Details" below) may question the consistency of the very foundations of the Everett's MWI.

{\bf Conclusion:} It seems that, in the case the decoherence is sufficient for the classical reality (as basically supposed to be correct in the MWI, i.e. the Multiverse theories), then one seems obliged to reject the Everett's (and alike) interpertations as physically incomplete, or possibly wrong.

\bigskip

{\bf Acknowledgements.} We are are looking forward to receiving the comments from (in alphabetical order)  H. Barnum, J. Halliwell, A. Yu Kamenshchik, H. P. Stapp,  M. Tegmark, F.J. Tipler, L. Vaidman, D Wallace...

\bigskip

TECHNICAL DETAILS

\bigskip

(a) the linear canonical transformations we have in mind read:

\begin{equation}
(x_{Si}, p_{Si}, X_{Ej}, P_{Ej}) \to (\xi_{S'k}, \pi_{S'k}, \Xi_{E'l}, \Pi_{E'l})
\end{equation}

\noindent for simplicity without any constraints for the transforming variables--thus preserving the number of the overall degrees of freedom of the World.

(b) the occurrence of decoherence for $S$ substantially (but not exclusively) depends on certain characteristics of the interaction $H_{S+E}$--e.g. of the so-called {\it separability} of the interaction. Virtually, the separability of interaction is  necessary for the occurrence of decoherence. The analysis refers equally to the exact (orthogonal) or to only approximate (approximatelly orthogonal) "pointer basis"; the later directly referring to both, the
so-called "back action" effect as well as to the (nonorthogonal) "preferred set of states".

(c) the linear canonical transformation give the different forms of the one and the same Hamiltonian:

\begin{equation}
H = H_S + H_E + H_{S+E} = H_{S'} + H_{E'} + H_{S'+E'}
\end{equation}

\noindent while, here, $H_{S'+E'} = 0$.

(d) the expression eq. (2) bears the clear physical meaning of the LCT in the context of the decoherence theory: in order to provide the decoherence effect for both decompositions, both the interaction terms, $H_{S+E}$ and $H_{S'+E'}$, should fulfill the separability criterion--that is probably the typical situation for the realistic models.

(e) nonequivalence requires some elaboration as given in the refs.:

[1] M. Dugi\' c, J. Jekni\' c, "What is 'system': some decoherence-theory arguments", Int. J. Theor. Phys. {\bf 45}, 2215 (2006)

[2] M. Dugi\' c, J. Jekni\' c-Dugi\' c, "What is 'system': the information-theoretic arguments", Int. J. Theor. Phys, {\bf 47}, 805 (2008)

\end{document}